\documentclass[aps,epsfig,floats,showpacs,twocolumn]{revtex4}
\usepackage{epsfig}
\begin{document}

\title{Inverse spin switch effects in Ferromagnet / Superconductor hybrids with strong ferromagnets}

\author{A.Yu.~Rusanov, S. Habraken and J.~Aarts}

\affiliation{Kamerlingh Onnes Laboratory, Leiden University, 2300
RA Leiden, The Netherlands}

\begin{abstract}
In F/S/F trilayers where the magnetization directions of the F layers can be
controlled separately, it has theoretically been predicted that the antiparallel
(AP) configuration can have a higher superconducting transition temperature $T_c$
than the parallel configuration. This is the so-called spin switch, which also
experimentally has been found for the case of weak ferromagnets.  Here we show that
strong ferromagnets yield the opposite effect. We study the transport properties of
F/S/F trilayers with F = Ni$_{0.80}$Fe$_{0.20}$ (Permalloy, Py) and S = Nb,
structured in strips of different sizes. Using two different thicknesses for the Py
layers, we can switch in a well-defined way between the AP- and P-configurations. In
the superconducting transition we find a clear increase of the resistance in the
AP-state. We ascribe this to enhanced reflection of spin-polarized quasiparticles at
the S/F interfaces which leads to a stronger suppression of superconductivity on the
S-side.
\end{abstract}

\date{\today}
 \maketitle \vskip 1truecm

One of the interesting phenomena which is currently searched for in hybrids of
superconductors (S) and ferromagnets (F) is the so-called superconducting
spin-switch effect. Basically, the effect can occur in F/S/F trilayers in which the
direction of the magnetization of one F-layer can be varied with respect to the
other. It was predicted some time ago that, in a current-in-plane (CIP) geometry,
and for a thickness of the S-layer $d_S$ of the order of its superconducting
coherence length $\xi_S$, the transition temperature $T_c^{AP}$ in the antiparallel
(AP) state is higher than the one in the parallel (P) state, $T_c^P$
\cite{buzdin99,balad01}. For particular choices of the different layer thicknesses
it should even be possible to find full re-entrant behavior, controlled with only a
small switching field \cite{tagirov99}. The effect is reminiscent of F/N/F spin
valves (N being a normal metal), with one important difference. In the normal metal
spin valves, the resistance is lowest in the parallel configuration, since in terms
of a two-spin-current model, it is determined by the spin channel with the smallest
resistance. In the superconducting case, the antiparallel configuration yields the
lowest (zero) resistance since the Cooper pair samples opposite exchange fields,
which are less pair-breaking than parallel fields.
\\
Full re-entrant behavior has not yet been observed in superconducting spin valves.
Two experiments were reported. Both were with a device consisting of weakly
ferromagnetic CuNi and superconducting Nb, and in both cases the reported $T_c^{AP}$
was only about 5~mK \cite{gu02a} or 2.5~mK \cite{potenza05} higher than $T_c^{P}$,
less than the width of the transition. The smallness of the effect is probably due
to the difficulty of producing highly transparent interfaces with these and similar
alloys. It is of importance to note that both theory and experiment are performed in
the limit of weak exchange field $h_{ex}$, with effectively one diffusion constant
for both spin species, very similar amounts of spin-up and spin-down particles, and
therefore weak spin polarization $P_s$. The effects of larger $h_{ex}$ or $P_s$ are
unknown and, at least for the CIP case, difficult to access by theory. \\
Here we want to show that high spin polarization actually leads to the opposite
effect. In trilayer combinations of ferromagnetic Permalloy (Ni$_{80}$Fe$_{20}$; Py)
with superconducting Nb, we find that, in the superconducting transition, switching
from the P to the AP configuration leads to an increase rather than a decrease of
the resistance when measured in a CIP geometry. The effect is strong; it is very
similar to recently reported findings on trilayers of ferromagnetic
La$_{0.7}$Ca$_{0.3}$MnO$_3$ (L), where $P_s$ is expected to be close to 100~\%) and
superconducting YBa$_2$Cu$_3$O$_7$ (Y) \cite{pena05}, and we offer a similar
explanation in terms of reflection of spin-polarized quasiparticles near the
interface \cite{takah99}. Where the conclusions in ref.~\cite{pena05} had to be
based on indirect evidence for the AP configuration, the use of Py allows to
demonstrate switching effects directly. Together, the experiments based on oxides
and on simple metals show that the effect is generic for the limit of high spin polarization. \\
Samples of Nb/Py were prepared by sputter deposition in an ultrahigh vacuum system.
Thick Nb films have a $T_c$ of 9.2~K, similar to the bulk. From the upper critical
fields we extract a value for the Ginzburg-Landau coherence length $\xi_{GL}(0)
\approx$ 13~nm. Using $\xi_{GL}(0) = 0.86\sqrt{\xi_0 \ell_N}$, with the BCS
coherence length $\xi_0 \approx$ 40~nm, this yields a value for the normal state
elastic mean free path $\ell_N \approx$ 5.5~nm. Samples were structured by e-beam
lithography in bridges of 0.5~mm $\times$ 4~mm ('large' samples) or in bridges of
3~$\mu$m $\times$ 20~$\mu$m ('small' samples). For both large and small samples we
used a design in which the contacts were included in the geometry and simple bars
with gold contacts in order to minimize problems with stray fields from contact pads
or arms. Ferromagnetic Py possesses a large spin polarization (45~\% \cite{Mood98}),
but also shows well-defined magnetization switching at low fields. Care was taken to
align the long axis of the bars with the easy axis of magnetization \^e$_{e}$, which
is induced by the residual magnetic fields in the sputtering machine. Magnetic
fields were applied in the plane of the sample, along the bars and therefore along
\^e$_{e}$. We also made use of the fact that the coercive field $H_c$ of the Py
layers depends on their morphology as well as on thickness. We consistently find
that a thicker Py layer deposited on the substrate has a lower value of $H_c$ than a
thinner layer deposited on top of the Nb layer. In this way it is quite possible to
have well-defined P- and AP-regimes. Typically, we used 50~nm for the Py bottom
layer and 20~nm for the Py top layer, which yields coercive or switching fields
around 1~mT - 2~mT (50~nm) and 8~mT - 10~mT (20~nm). For the Nb layer the smallest
thickness $d_{Nb}$ was around 25~nm, which yields transition temperatures around 4~K
and is already close to the critical thickness $d_{cr}$ for the trilayer.\\
\begin{figure}
{\includegraphics[width=6cm]{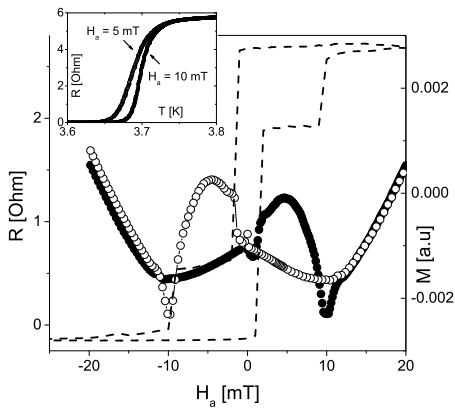}} \caption{Resistance $R$ versus
applied field $H_a$ of a 'large' sample $s$/Py(50)/Nb(25)/Py(20) at 3.66~K; filled
symbols : forward sweep, open symbols : backward sweep. Dashed : magnetization $M$
versus $H_a$ at 5~K of a similar but unstructured sample. Insert : $R$ versus
temperature $T$ at 5~mT in the P-state and at 10~mT in the domain state.}
\label{inv-switch-fig1}
\end{figure}
Fig.~\ref{inv-switch-fig1} shows a compilation of different measurements on a sample
$s$/Py(50)/Nb(25)/Py(20) (with $s$ denoting the substrate and the numbers the layer
thickness in nm ). The magnetization was measured at 5~K on the unstructured sample
by SQUID magnetometry. The switching of the layers is well defined and from the
magnitude of the jumps it can be seen that the 50~nm layer switches at $\pm H_{c,50}
\approx$~1.5~mT, and the 20~nm layer at $\pm H_{c,20} \approx$~9.5~mT, leaving a
large field range for the AP-state. For transport measurements, the sample was
structured as a large bar with gold contacts and showed a resistive transition
around 3.7~K with a width $\Delta_{tr}$ of 100~mK. The resistance $R$ was measured
as function of the in-plane field $H_a$ at a temperature of 3.66~K, as shown in
Fig.~\ref{inv-switch-fig1}. Starting at high fields, $R$ decreases until $+H_{c,20}$
where it starts to rise slowly. At $-H_{c,50}$ a small but clear upward jump occurs.
This is the field where the alignment of the Py layers becomes AP. In this regime
$R$ rises further to a peak, followed by a steep decrease to a dip at around
$-H_{c,20}$. Now the sample is in the P state and $R$ starts to rise slowly again.
The behavior is mirrored in increasing fields. The strong peaks in the resistance
therefore appear to be connected to the AP alignment, just as in the case of the
L/Y/L trilayers of ref~\cite{pena05}. The dips at $\pm H_{c,20}$ are well known and
are produced by the magnetic domains which occur around the Py coercive field in the
20~nm layer. They can also be found in Py/Nb bilayers, and are due to a lower
averaged exchange field sensed by the Cooper pair, as we have demonstrated
previously \cite{rusanov04}. For the 50~nm layer they also should be present, and a
small dip is actually observed in backward sweep at +1.5~mT, but it is masked by the
increase of resistance resulting from the P $\rightarrow$ AP switch. The effect of
the domain state on $T_c$ is shown in the insert, where R(T) is given at two
different fields; at 5~mT coming from high field (sample in the P-state), and at
10~mT coming from low fields (sample in the domain state). The difference in the
temperature where zero resistance is reached is at most 30~mK, similar to the
earlier findings. Finally, the rise in $R$ at $+H_{c,20}$ when coming from high
fields may appear puzzling, since no switching takes place at this field. Domain
formation, however, already does set in : close inspection of M(H) shows that
decrease already starts before $H_{c,50}$ is
reached, and this should lead to small amounts of AP orientations. \\
The data therefore suggest that the AP state shows larger resistance than the P
state, but for the large sample the behavior is sluggish because of domain effects.
Next we consider some much smaller samples.
\begin{figure}
{\includegraphics[width=6cm]{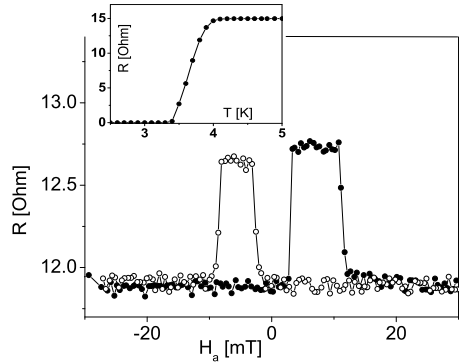}} \caption{Resistance $R$ versus
applied field $H_a$ of a 3~$\mu$m $\times$ 20~$\mu$m bridge (contacts included) of a
sample $s$/Py(50)/Nb(26)/Py(20) at 3.8~K; filled symbols : forward sweep, open
symbols : backward sweep. Insert : $R$ versus temperature $T$.}
\label{inv-switch-fig2}
\end{figure}
Fig.~\ref{inv-switch-fig2} shows data on a sample $s$/Py(50)/Nb(26)/Py(20) with a
bridge of 3~$\mu$m $\times$ 20~$\mu$m and contacts included. The transition (see
inset) is quite broad, $\Delta_{tr} \approx$ 600~mK and the measurement is taken at
3.80~K, close to the onset at 4~K. After correcting for a small offset field, the
values for $\pm H_{c,50}$ and $\pm H_{c,20}$ are 2.7~mT and 10~mT, respectively. The
switching behavior is now perfectly well defined. Also noticeable is the absence of
the dips which we ascribed to the domains. This is again in agreement with our
earlier observations and due to the fact that no stable domain state is formed in
these small samples during the switching \cite{rusanov04}. Samples with similar
thicknesses show the same behavior, although the switching is not always a perfect
one-step process; sometimes, several steps (both up and down) can be seen. This may
not be surprising, since the properties of Py, and the direction of the easy axis
are very sensitive to the preparation conditions. Next we increase $d_{Nb}$. The
transition width now gradually decreases. Fig.~\ref{inv-switch-fig3} shows data on a
sample $s$/Py(50)/Nb(60)/Py(20) with $\Delta_{tr} \approx$ 100~mK and a normal state
resistance of 9.89~$\Omega$. In this case, the sample was a simple bar of 3~$\mu$m
$\times$ 20~$\mu$m with Au contacts, which we show to make clear that the effect can
be found with different contact geometries. The behavior of $R(H_a)$ is shown at T
=~7.46~K, halfway the transition, and at T =~7.40~K, close to the bottom. The
switching behavior is still sharp and clear, quite similar to the previous sample.
Note that the size of the resistance variation has become much smaller at the lower
temperature. For samples with $d_{Nb}$ =~80, 90~nm and $\Delta_{tr} \leq$ 50~mK, the
effect became very small.
\begin{figure}
{\includegraphics[width=6cm]{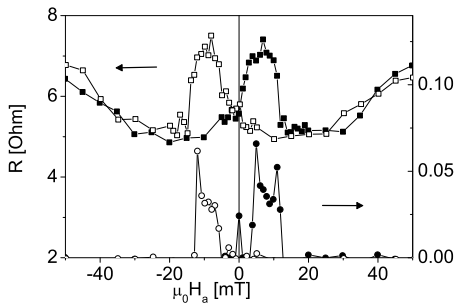}} \caption{Resistance $R$ versus
applied field $H_a$ of a 2~$\mu$m $\times$ 20~$\mu$m bridge of a sample
$s$/Py(50)/Nb(60)/Py(20) at 7.46~K (squares) and at 7.40~K (circles); filled symbols
: forward sweep, open symbols : backward sweep. } \label{inv-switch-fig3}
\end{figure}
Above $T_c$,$R(H_a)$ of the samples only shows tiny dips around $\pm H_{c,50}$,
which are due to the anisotropic magnetoresistance (AMR) effect for configurations
where the current is parallel to the applied magnetic field \cite{o'hand00}. \\
The observed effects are therefore due to the onset of the superconducting state.
For all geometries, in the resistive transition, the AP state has a higher
resistance than the P state, which is opposite to the effects predicted
\cite{buzdin99,tagirov99} and observed for weak ferromagnets \cite{gu02a,potenza05},
but similar to the observations of ref~\cite{pena05}. In terms of the
magnetoresistance $MR$~= $\Delta R$ = $(R_{AP}-R_P)/R_P$, the effect becomes quite
large due to the decreasing value of $R$, but for the physics it is more relevant to
note that in terms of the normal-state resistance of the samples, it is a fraction
of the order of 5~\% - 10~\%, at least in the upper half of the transition. It is
then of interest to compare the results of these CIP measurements to the results of
current-perpendicular to-plane (CPP) measurements on stacks of Py/Nb/Py, performed
by Gu {\it et al.} \cite{gu02b}, for different thicknesses $d_{Nb}$ of the Nb layer
in a range between 30~nm and 100~nm. In the CPP-case, a small (1~\%) positive
($R_{AP} > R_P$) effect was present in the normal state, which persisted below
$T_c$. However, it was found to decrease with decreasing temperature, which is
different from our CIP data which show an initial increase of MR (from 0 in the
normal state). A basic explanation of the CPP data was given in terms of the
diffusion of spin-polarized quasiparticles (qp) with energies $E_{qp}$ below the
gap. The MR effect in the normal state is due to the standard mechanism of increased
spin scattering in the AP configuration. This becomes smaller through spin memory
loss controlled by the spin diffusion length $\ell_{sd}$, but when the intermediate
layer is a superconductor, where spin is carried only by quasiparticles and not by
Cooper pairs, MR will also decrease due to the loss of quasiparticles to the
condensate, which is controlled by the qp diffusion length $\xi_{qp}$ \cite{note1}.
From the temperature dependence of $\Delta R$, a value of $\xi_{qp} \approx$ 16.5~nm
was found, very close to $\xi_{GL}(0) \approx$ 13~nm of the Nb. This is actually not
surprising. It was already shown by Blonder, Tinkham and Klapwijk that the
characteristic decay of evanescent quasiparticles with energies inside the gap (the
current-to-supercurrent conversion length) is given by 1.22$\xi_{GL}(T)$
\cite{BTK82}. The number also shows that spin loss by spin scattering only plays a
minor role, since $\xi_{GL}(0) << \ell_{sd}$ $\approx$ 50~nm \cite{gu02b}. This
analysis was confirmed in recent theoretical work by Yamashita {\it et al.}
\cite{yama03}, who considered Andreev reflections and direct transmission of
spin-polarized quasiparticles in F/S/F systems. They found slightly larger values
for $\xi_{qp}$, and also that $\xi_{qp}$ decreases with increasing $d_{Nb}$. This
was ascribed to the proximity effect in this all-metal system, which suppresses the
average gap as long as $d_{Nb}$ is
comparable to $\xi_{GL}$. \\
Translating this description to our CIP case, the first thing to note is the
differences in MR in the normal state. This is easily explained. In CPP, for
diffusive systems, the dependence of $\Delta R / R$ on the spacer thickness $d_N$ is
$\propto e^{-d_N / 2 \ell_{sd}}$ \cite{valet93}. If MR effects are present in the
Py/Nb/Py system, they can be witnessed for the range of thicknesses used in the CPP
experiment. On the other hand, in CIP the attenuation is $\propto e^{-d_N /
\ell_N}$, with $\ell_N$ the elastic mean free path of the normal metal
\cite{sperio93,coeh00}. For our Nb, we estimated before that $\ell_N \approx$
5.5~nm, much smaller than the spacer thickness of 25~nm. So, if MR effects are found
in CPP, they will not be observed in CIP in the same thickness range. This changes
in the superconducting transition, since then quasiparticles appear with a much
longer range because of the divergence close to $T_c$. We now offer a similar line
of reasoning as ref.~\cite{pena05}. Current is flowing in the plane of the films,
there is no voltage difference perpendicular to the layers, so electrons scattering
out of the F layer turn into low-energy spin-polarized quasiparticles, which can
diffuse across the S layer. In the AP configuration, a larger number experiences
reflection at the other interface than in the P configuration. Although the spin
determines this reflection process, the result should probably not be called spin
accumulation since there is no net charge or spin transport through the interface
\cite{valet93}, in contrast to the CPP case. The reflection however also leads to a
larger number of quasiparticles on the S-side for the AP case than for the P case,
and this translates (selfconsistently) into a gap suppression on the S-side. This
gap suppression is observed as an increased resistance.\\
Based on the magnitude of the switching, we can estimate the size of the layer where
this takes place for temperatures close to $T_c$, at the top of the transition,
where the order parameter profile is still flat. In the normal state, the specific
resistances are $\rho_{Nb}$ = 7.5~$\mu \Omega cm$ for the Nb and $\rho_{Py}$ =
30~$\mu \Omega cm$ for the Py. The contribution to the resistance of the Nb layer in
a sample $s$/Py(50)/Nb(25)/Py(20) is therefore 40~\%. Assuming that the full
variation in $\Delta R / R_N \approx$ 10~\% is due to extra resistance in the Nb
layer, this corresponds to 25~\% of the Nb layer, which is 6~nm, or 3~nm on each
side. This rough estimate suggests a suppressed layer with a thickness of the order
of $\ell_N$. Finding the exact dependence of $\Delta R$ on temperature and thickness
needs more elaborate modelling. The number of quasiparticles decreases with
decreasing temperature and increasing gap size, and this partly explains the
smallness of $\Delta R$ for the sample with $d_{Nb}$ = 50~nm near the bottom of the
transition, but it also a function of thickness. Close to $d_{cr}$, which is around
2$\xi_{GL}(0)$, the order parameter stays flat, but with increasing thickness it
will grow more strongly in the middle of the layer and act as a bottleneck for the
evanescent quasiparticles, since their accessible energy range becomes smaller. This
should lead to a decreasing transition width, as witnessed, but also to different
behavior of $\Delta R$ as function of reduced temperature $T/T_c$. Also the effect
of the spin diffusion length will start to play a role around $d_{Nb} \approx$ 50~nm
$\approx \ell_{sd}$ , but because of the above argument it cannot be claimed that
the loss of MR effects for our thickest samples is simply due to the loss of spin. \\
In conclusion, our observations put a new perspective on the feasibility of the
superconducting spin switch. Close to $T_c$, spin switch effects can be found with
weak ferromagnets but the difficulty of obtaining highly transparent interfaces when
using strongly disordered alloys may preclude full switching. Increasing the
polarization, however, leads to a competing effect, namely the increased quenching
of the superconductivity when the AP configuration reflects more spin-polarized
quasiparticles back into the superconductor than the P-configuration. Note that both
mechanisms are not mutually exclusive since one depends on the quasiparticles and
the other on the Cooper pairs. We also have made clear the differences between CIP
and CPP experiments. Finally, our observations should be of importance for the
reproducibility and interpretation of data from S/F multilayers with strong magnets
in general. In many reported experiments on $T_c$-variations, the magnetization
state of the sample is undefined, and in particular re-entrant effects close to the
critical thickness might be affected by the domain state of the sample. \\

We thank J. Santamaria, C. Bell and M. Flokstra for useful discussions, and M.
Hesselberth for help in the sample preparation. This work is part of the research
program of the "Stichting voor Fundamenteel Onderzoek der Materie (FOM)", which is
financially supported by NWO. The ESF-programs 'Pishift' and 'Thiox' are
acknowledged for providing an invaluable forum for discussing the preliminary
results.


\begin{thebibliography}{10}
\bibitem{note1} In ref.~\cite{gu02b} the notation $\ell_{qp}$ was used for what
was called the quasiparticle penetration depth. We prefer $\xi_{qp}$ in order to
stress the superconducting nature of the length scale, which is reflected in its
divergence at $T_c$.
\end{thebibliography}
\end{document}